\documentclass[journal]{IEEEtran}

\ifCLASSINFOpdf
\else
   \usepackage[dvips]{graphicx}
   
\fi
\usepackage{url}

\usepackage{graphicx}
\usepackage{amssymb}
\usepackage{amsmath}
\usepackage{xcolor}
\usepackage{balance}

\def\minwrt[#1]{\underset{#1}{\mathrm{minimize }}}
\def\minwrtshort[#1]{\underset{#1}{\mathrm{min. }}}

\def\argminwrt[#1]{\underset{#1}{\mathrm{arg min }}}

\def\br{{\mathbf r}}
\def\bn{{\mathbf n}}
\def\bg{{\mathbf g}}
\def\bPhi{{\boldsymbol{\Phi}}}
\def\bDelta{{\boldsymbol{\Delta}}}
\def\bmu{{\boldsymbol{\mu}}}
\def\bnu{{\boldsymbol{\nu}}}

\def\RR{\mathbb{R}}
\def\CC{\mathbb{C}}
\def\TT{\mathbb{T}}
\def\mpos{\mathcal{M}_+}

\def\soundspeed{\rho}
\def\groundcost{c}
\def\ot{\tilde{T}}
\def\otvec{T}

\newcommand{\abs}[1]{\left|#1\right|}

\begin{document}

\title{Sound Field Estimation Using Optimal Transport Barycenters in the Presence of Phase Errors}

\author{Yuyang Liu, Johan Karlsson \IEEEmembership{Senior member, IEEE}, and Filip Elvander, \IEEEmembership{Member, IEEE}
\thanks{This research was supported in part by the Research Council of Finland
(decision number 362787) and in part by the Swedish Research Council (VR) under grant 2020-03454.}
\thanks{Yuyang Liu and Filip Elvander are with the Department of Information and Communications
Engineering, Aalto University, Finland (e-mail:
yuyang.liu@aalto.fi, filip.elvander@aalto.fi).}
\thanks{Johan Karlsson is with the Department of Mathematics, KTH Royal Institute of Technology, Sweden (e-mail:johan.karlsson@math.kth.se).}}

\markboth{}
{Shell \MakeLowercase{\textit{et al.}}: Bare Demo of IEEEtran.cls for IEEE Journals}
\maketitle

\begin{abstract}
This study introduces a novel approach for estimating plane-wave coefficients in sound field reconstruction, specifically addressing challenges posed by error-in-variable phase perturbations. Such systematic errors typically arise from sensor mis-calibration, including uncertainties in sensor positions and response characteristics, leading to measurement-induced phase shifts in plane wave coefficients. Traditional methods often result in biased estimates or non-convex solutions. To overcome these issues, we propose an \textit{optimal transport} (OT) framework. This framework operates on a set of \textit{lifted} non-negative measures that correspond to observation-dependent shifted coefficients relative to the unperturbed ones. By applying OT, the supports of the measures are transported toward an optimal average in the phase space, effectively morphing them into an indistinguishable state. This optimal average, known as barycenter, is linked to the estimated plane-wave coefficients using the same lifting rule. The framework addresses the ill-posed nature of the problem, due to the large number of plane waves, by adding a constant to the ground cost, ensuring the sparsity of the transport matrix. Convex consistency of the solution is maintained. Simulation results confirm that our proposed method provides more accurate coefficient estimations compared to baseline approaches in scenarios with both additive noise and phase perturbations.
\end{abstract}

\begin{IEEEkeywords}
Sound field estimation, phase error, lifting, optimal transport barycenter
\end{IEEEkeywords}

\IEEEpeerreviewmaketitle

\section{Introduction}
\label{sec:intro}


Modeling and estimation of sound fields, i.e., the complex pressure over space, is essential for applications such as auralization ~\cite{nicol19993d,hulsebos2002improved}, noise cancellation ~\cite{koyama2021spatial,samarasinghe2016noisecontrol}, and sound zone control \cite{JesperSZC,BetlehemPSZ}. For example, in sound zone control, the design of control filters depends on having access to an accurate description of the acoustic environment \cite{koyamaSFC}. Sound field models typically fall in two broad categories: geometrical models and physical models \cite{antonello2017room,betlehem2005theory}. Geometrical models estimate the locations and contributions of the reflected waves \cite{crocco2015room}, whereas physical models take their cue from differential equations describing sound propagation \cite{mignot2013low}. In practice, the parameters of a sound field model are typically inferred by fitting the model to a finite set of microphone measurements. To avoid intractable inference problem over general function spaces, one may add additional structure and regularization, as in so-called kernel-ridge regression approaches \cite{Davidkernel,koyamaKernel}, or in finite discretization schemes using, e.g., truncated plane-wave decompositions \cite{samarasinghe2015efficient}. In the latter case, regularization in the form of sparsity-promoting penalities \cite{georgios2010lasso}, Tikhonov regularization \cite{wu2000reconstruction}, or Bayesian approaches \cite{sundstrom2025boundary}, may be employed to render the inference problem well-posed as the number of vectors in the expansion typically is considerably larger than the number of measurements \cite{verburg2018reconstruction}. 

However, expansion approaches are dependent on well-calibrated setups as the expansion vectors employed depend on, e.g., sensor positions and response characteristics. As a consequence, in the presence of calibration errors, sensor measurements may be subject to error-in-variables-type perturbations in addition to additive sensor noise typically considered \cite{royster2003sound,buck2002aspects}. In this work, we focus on the case where mis-calibration manifests as perturbations of the phase of the complex-valued coefficients in plane-wave expansion models. Such errors may arise due to uncertainty in the sensor positions as changes in time-delay correspond to phase-shifts. However, introducing phase-errors in the plane-wave decomposition models renders the problem of estimating the expansion coefficients biased or non-convex. To address this, we propose a lifting approach where we represent complex-valued coefficients using non-negative measures on the unit circle. In this representation, phase-shifts can be understood as perturbations of the corresponding measures. This allows us to leverage the concept of optimal transport in order to quantify the discrepancy between complex scalars of the same modulus. 

Optimal transport is a mathematical framework that allows comparing and quantifying distances between non-negative measures based on minimal displacements of "mass" \cite{villani2009optimal}. Recently, this concept has found application in signal processing and machine learning \cite{gabriel2019computational,Kolouri2017omtApp}, as well as in room acoustics in the context of room impulse response estimation and modeling~\cite{RIRInterDavid,RIRSimuInfe,RIRotBarycenter}.
%
In our setting, we model the effective plane-wave coefficients at the different sensors as phase-perturbed versions of nominal coefficients at a virtual reference point. The corresponding non-negative measure explaining these coefficients can then be understood as an OT barycenter, or generalized mean, which we can compute using convex optimization techniques. This is reminiscent of, but distinct from, earlier works on OT for power spectra based on observed Fourier coefficients \cite{PSMOT,WassersteinFourierD,OMToeplitzInterExtra}.
%
%
%
Furthermore, the proposed formulation naturally induces sparse solutions, promoting parsimonious representations where only a small number of plane-waves is active in the expansion.

\section{Signal model}
\label{ProblemFormulation}
Consider a set of $Q\in \mathbb{N}$ microphones located in a region $\Omega \subseteq \mathbb{R}^3$. Then, under the assumption of $\Omega$ being source-free, the frequency-domain (complex) sound-field $p(\br)$ satisfies the Helmholtz equation
\begin{equation*}
  \left(\nabla^2 + k^2\right)p(\br) = 0, 
\end{equation*} 
where $\br \in \Omega$ is observation position, $k = 2\pi f / \soundspeed$ is the wave-number, with $f$ being the frequency and $\soundspeed$ the speed of sound~\cite{williams1999fourier}. Such sound-fields allow for decomposition into series of plane-waves, i.e., components of the form $e^{-i{k\bn}^T\br}$ where $\bn$ is a unit-length direction vector \cite{kuttruff2016room}. In practice, the sound-field is often approximated as a finite sum \cite{antonello2017room}
\begin{align*}
    p(\br) \approx \sum_{\ell=1}^L \Phi_\ell(\br) \;,\; \Phi_\ell(\br) = \alpha_\ell e^{-i{k\bn_\ell}^T\br}
\end{align*}
where $\alpha_\ell \in \CC$ are complex amplitudes and $\bn_\ell$ are the directions of the impinging waves. Define the vector $\bPhi = \begin{bmatrix} \alpha_1 & \ldots & \alpha_L \end{bmatrix}^T \in \CC^L$ and the steering vector function $g: \Omega \to \CC^L$ as
\begin{align*}
    g(\br) = \begin{bmatrix} e^{-i{k\bn_1}^T\br} & \ldots & e^{i{k\bn_L}^T\br}  \end{bmatrix}^T.
\end{align*}
Then, the complex pressure at any point can be computed as
\begin{align*}
    p(\br) = \langle g(\br), \bPhi \rangle.
\end{align*}
In particular, the pressure measured by microphone $q$ is 
\begin{align*}
    p^{(q)} \triangleq p(\br^{(q)}) = \langle \bg^{(q)}, \bPhi \rangle \;,\; \bg^{(q)} = g(\br^{(q)}),
\end{align*}
where $\br^{(q)}$ is the microphone position. In practice, the cardinality $L$ of the set of directions $\bn_\ell$ is typically significantly larger than the number of measurements, $Q$, rendering the problem of recovering $\bPhi$ from (noisy) observations of $p^{(q)}$ ill-posed. To remedy this, sparse estimation techniques from the compressed sensing literature has successfully been employed, typically formulated as regularized versions of least-squares estimation \cite{verburg2018reconstruction}. However, in addition to additive noise, the microphone measurements may also be subject to calibration errors. In particular, errors arising from, e.g., position uncertainties, alignment errors, or scattering effects, may manifest themselves as systematic perturbations in the measurement phase. That is, there will be vectors $\bDelta^{(q)} \in \RR^L$ such that the (noise-free) measurements correspond to $\langle \bg^{(q)}\odot e^{i\bDelta^{(q)}}, \bPhi \rangle$, $q = 1,\ldots,Q$, where $\odot$ is the Hadamard product and exponentiation is elementwise. The resulting noisy measurement model is
\begin{align}\label{eq:noisy_measurements}
    \tilde{p}^{(q)} =  \langle \bg^{(q)}\odot e^{i\bDelta^{(q)}}, \bPhi \rangle + \epsilon^{(q)},
\end{align}
where $\epsilon^{(q)}$ denotes the additive noise. In this scenario, employing standard sparse estimation techniques for recovering $\bPhi$ directly from the set $\{\tilde{p}^{(q)} \}_{q = 1}^Q$ will result in biased estimates. Although one can consider methods such as total least squares \cite{golub1980analysis}, these typically do not take into the account that the error-in-variables correspond to perturbations on the complex unit circle. Furthermore, the resulting estimation problems are non-convex and not easily extended. Herein, we aim to address the problem of recovering $\bPhi$ under both phase perturbations and additive noise using a lifting procedure, allowing us to arrive at estimator formulated as a convex optimization problem. In particular, we will lift the complex-valued plane wave coefficients to the space of non-negative measures on the circle, allowing us to invoke the machinery of optimal transport to solve the estimation problem.
\begin{figure*}[t]   
    \centering
    \includegraphics[width=1\textwidth,trim=120 10 120 0,clip]{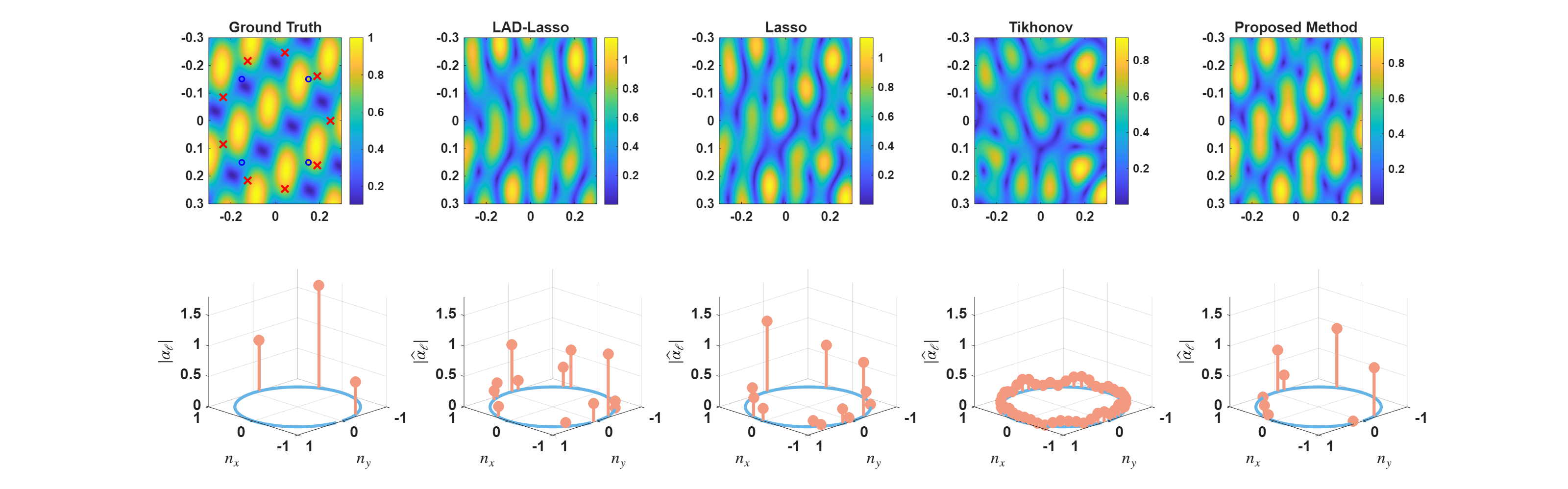}
    \caption{Top row: Magnitude of the sound field generated by three plane waves.
Bottom row: Magnitude of the plane-wave coefficients, where $\bn=(n_x,n_y)$ represents the unit-length direction vector. Non-zero positions indicate the incident wave direction. Reconstructions using LAD-Lasso, Lasso, Tikhonov, and the proposed method, based on nine measurement points, are compared to the ground truth. Measurement sensor locations are marked with red crosses in the ground truth panel, and cross-validation points are shown in blue.}
    \label{NR:SFR}
\end{figure*}
\section{Proposed method}

\subsection{Lifting}

We note that the noise-free measurements can be written as
\begin{align*}
    \langle \bg^{(q)}\odot e^{i\bDelta^{(q)}}, \bPhi \rangle = \langle \bg^{(q)} , e^{-i\bDelta^{(q)}}\odot\bPhi \rangle = \langle \bg^{(q)} , \bPhi^{(q)} \rangle
\end{align*}
where $\bPhi^{(q)} \triangleq e^{-i\bDelta^{(q)}}\odot\bPhi$, $q =1,\ldots,Q$, are set of phase-perturbed coefficients. Then, we have
\begin{align*}
    \bPhi^{(q)} = \abs{\bPhi}\odot e^{i(\angle\bPhi - \bDelta^{(q)})}
\end{align*}
where $\abs{\cdot}$ and $\angle(\cdot)$ denotes elementwise absolute value and complex angle, respectively. Define $\TT \triangleq [-\pi,\pi)$, which we identify with the unit circle, and let $\mpos(\TT)$ and $\mpos^L(\TT)$ be the set of non-negative measures and $L$-vector valued non-negative measures on $\TT$. Then, we may associate $\bPhi^{(q)}$ with element $\bmu^{(q)} \in \mpos^L(\TT)$. To see this, construct for the $\ell$th element of $\bPhi^{(q)}$ the measure $\mu^{(q)}_\ell \in \mpos(\TT)$ as
\begin{align*}
    \mu^{(q)}_\ell(\psi) = \abs{\alpha_\ell} \delta(\psi - (\angle \alpha_\ell - \Delta_\ell^{(q)})),
\end{align*}
where $\delta$ is the Dirac delta. We may recover the $\ell$th coefficient of $\bPhi^{(q)}$ as 
\begin{align}\label{eq:first_Fourier_coeff}
\alpha_\ell \cdot e^{-i\Delta_\ell^{(q)}} = \int_{\TT}e^{i\psi}d\mu_\ell^{(q)}(\psi).
\end{align}
Note here that $\mu_\ell^{(q)}$ assigns all its mass to the point $\angle \alpha_\ell - \Delta_\ell^{(q)}$. With this construction, we have that
\begin{align*}
    \bmu^{(q)} \triangleq \begin{bmatrix}
        \mu_1^{(q)} &  \mu_2^{(q)} & \ldots & \mu_L^{(q)}
    \end{bmatrix}^T
\end{align*}
represents $\bPhi^{(q)}$ in the sense that
\begin{align*}
    \bPhi^{(q)} = \int_\TT e^{i\psi} d\bmu^{(q)}(\psi).
\end{align*}
Furthermore, it follows directly that
\begin{align*}
    \int_\TT d\bmu^{(q)}(\psi) = \abs{\bPhi}, \; q = 1,\ldots,Q.
\end{align*}
That is, all measures $\bmu^{(q)}$ have the same total mass, corresponding to the magnitudes of the plane-waves. In other words, $\mu_\ell^{(q)}$ is a measure whose zeroth and first Fourier coefficients are $\abs{\alpha_\ell}$ and $\alpha_\ell \cdot e^{-i\Delta_\ell^{(q)}}$, respectively. They differ only in their support, which mirrors the effect of the phase perturbations $\bDelta^{(q)}$. We may now re-phrase the measurement model as
\begin{align}\label{eq:noisy_measurements_mu}
    \tilde{p}^{(q)} =  \left\langle \bg^{(q)}, \int_\TT e^{i\psi}d\bmu^{(q)}(\psi) \right\rangle + \epsilon^{(q)}.
\end{align}
 Then, under the assumption that the perturbations are relatively small, we can understand the problem of phase correction as that of morphing the set of distributions $\bmu^{(q)}$ as to become indistinguishable, while preserving the total mass. To find the minimal, or most efficient, morphing may then be phrased as an optimal transport problem. Specifically, we will formulate the recovery of $\bPhi$ as finding a corresponding measure $\bmu^{(0)} \in \mpos^L(\TT)$, where $\bmu^{(0)}$ is an optimal transport barycenter.
\subsection{Optimal Transport Barycenter}\label{OTBarycenter}
Consider two measures $\mu,\nu \in \mpos(\TT)$ and let $\groundcost:\TT\times\TT\to\RR_+$ be a $2\pi$-periodic and continuous so-called ground cost. Then, we may quantify the discrepancy between $\mu$ and $\nu$ by means of a Monge-Kantorovich problem of optimal transport \cite{villani2009optimal} according to
\begin{equation}\label{eq:basic_ot}
\begin{aligned}
    \ot(\mu,\nu)\!
    \!= \min_{m\in\mpos(\TT\times\TT)}& \int_{\TT\times\TT}\groundcost(\psi_1,\psi_2) d m(\psi_1,\psi_2)
    \\\mbox{subject to }& \int_{\TT}\!\!  dm(\cdot,\psi_2) =\mu, \int_{\TT}\!\! dm(\psi_1,\cdot)  = \nu.
\end{aligned}
\end{equation}
Here, the so-called transport plan $m$ describes how mass is moved between $\mu$ and $\nu$, with the constraints ensuring that all mass is accounted for. As the cost of mass movement is determined by $\groundcost$, the minimal objective, i.e., $\ot(\mu,\nu)$ corresponds to the most efficient way of morphing, or perturbing, $\mu$ as to become identical to $\nu$. Herein, we will pick the ground cost as $\groundcost(\psi_1,\psi_2) = \abs{e^{i\psi_1} - e^{i\psi_2}}^2 + \gamma$, where $\gamma > 0$. We will elaborate on this choice in Section~\ref{sec:estimator}. Building on this, one may for a set of measures $\mu^{(q)} \in \mpos(\TT)$ define their OT barycenter, or generalized average, as
\begin{align*}
    \mu^{(0)} = \argminwrt[\mu \in \mpos(\TT)] \;\; \frac{1}{Q}\sum_{q=1}^Q \ot(\mu, \mu^{(q)}).
\end{align*}
That is, $\mu^{(0)}$ minimizes the average OT distance to the set $\{ \mu^{(q)}\}_{q=1}^Q$. In the context of $\mu^{(q)}$ modeling a phase perturbation of the form \eqref{eq:first_Fourier_coeff}, the barycenter $\mu^{(0)}$ can be interpreted as a type of phase-averaging. Furthermore, an estimate of a corresponding plane-wave coefficient $\alpha \in \CC$ can be constructed as $\hat{\alpha} = \int_\TT e^{i\psi} d\mu^{0}(\psi)$. We may directly extend this to vector-valued measures according to 
\begin{align*}
    \bmu^{(0)} = \argminwrt[\bmu \in \mpos^L(\TT)] \;\; \frac{1}{Q}\sum_{q=1}^Q \otvec(\bmu, \bmu^{(q)}),
\end{align*}
where for $\bmu, \bnu \!\in\! \mpos^L(\TT)$, we define $\otvec(\bmu,\bnu)\!\!=\!\!\sum_{l = 1}^L \ot(\mu_l,\nu_l)$,
%
%
i.e., transport is performed between the different components $(\mu_\ell,\nu_\ell)$ of $\bmu$ and $\bnu$. It may be verified $\ot$ is jointly convex in its arguments, implying that $\otvec$ is also jointly convex.
\begin{figure*}[t]
\centering

\begin{minipage}[t]{0.33\textwidth}
    \centering
    \includegraphics[width=\linewidth]{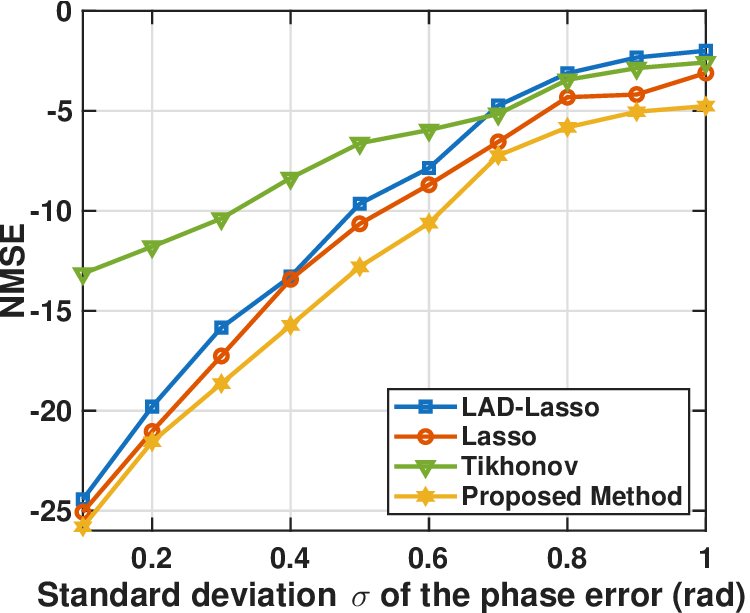}
    \label{fig:vary_phase_error}
\end{minipage}\hfill
\begin{minipage}[t]{0.316\textwidth}
    \centering
    \includegraphics[width=\linewidth]{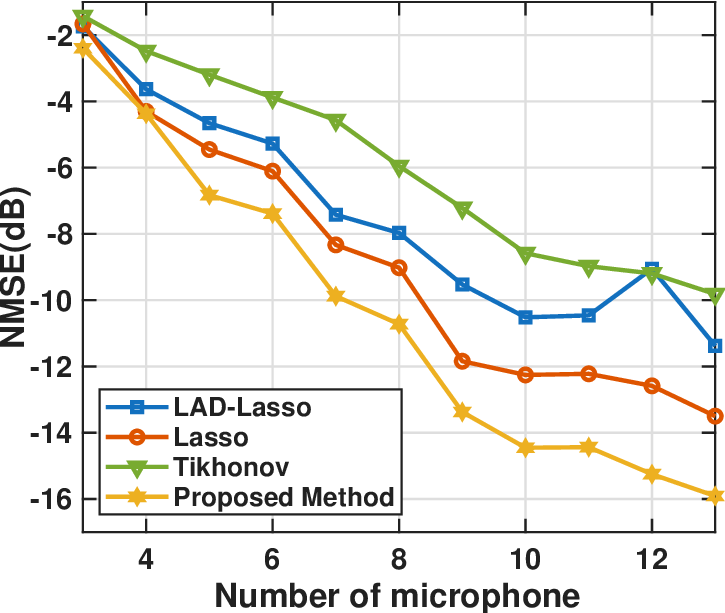}
    \label{fig:vary_nbr_mics}
\end{minipage}\hfill
\begin{minipage}[t]{0.31\textwidth}
    \centering
    \includegraphics[width=\linewidth]{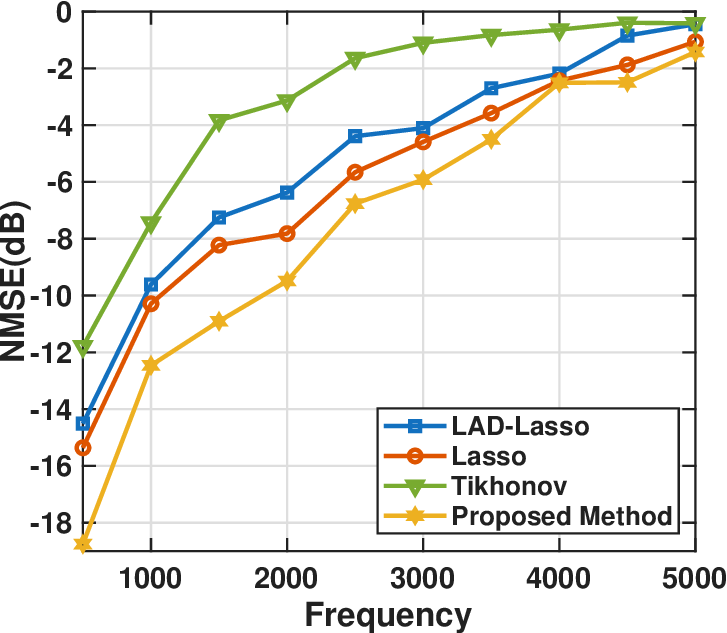}
    \label{fig:vary_F}
\end{minipage}
    \caption{NMSE as a function of system parameters based on 30 Monte Carlo experiments. Left: NMSE versus the number of microphones; Middle: NMSE versus frequency; Right: NMSE versus the standard deviation of the phase error.}
    \label{fig:vary_parameters}
\end{figure*}
\subsection{Resulting estimator} \label{sec:estimator}
%
We can now define our estimator of $\bPhi$ as
\begin{align}\label{eq:ot_est}
    \hat{\bPhi} = \int_{\TT} e^{i\psi} d\bmu^{(0)}(\psi),
\end{align}
where $\bmu^{(0)}$ solves the inverse barycenter problem
\begin{equation*}
\begin{aligned}
    \minwrtshort[\bmu^{(0)}, \bmu^{(q)}] & \;\sum_{q = 1}^Q \otvec(\bmu^{(0)},\bmu^{(q)})
     \!+\!\eta  \left|\!\left\langle\! \bg^{(q)},\!\! \int_{\TT}\!\!\! e^{i\psi}\!\,\mathrm{d}\bmu^{(q)}(\psi)\right\rangle
\!\!-\! \tilde{p}^{(q)}\right|^2\!,
\end{aligned}
\end{equation*}
and where $\eta > 0$ is a user-defined parameter weighting the relative importance between the transport cost and data fit. The choice of the $\ell_2$ penalty may be interpreted as assuming that the noise $\epsilon^{(q)}$ is independent Gaussian. Note here that we simultaneously estimate the phase-perturbed plane-wave coefficients at the microphones $q$, corresponding to the first Fourier coefficients of the respective $\bmu^{(q)}$, and the nominal, unperturbed plane-wave coefficients as modeled by $\bmu^{(0)}$. The assumption that the phase-perturbations are small is reflected by the transport term, promoting distributions over $\TT$ that are similar in the transport sense.
Furthermore, in order to exactly model phase-perturbations, the component measures $\mu^{(0)}_\ell, \mu^{(q)}_\ell$ should ideally consist of a single Dirac measure each. In order to promote such solutions, the inclusion of $\gamma > 0$ in the ground cost $\groundcost$ is critical. In particular, $\gamma$ serves as a penalty on the total mass, which may be noted to not be directly observable from the measurements \eqref{eq:noisy_measurements_mu}. Then, as the barycenter problem corresponds to a linear program, solutions will consist of exactly such Dirac measures for large enough $\gamma$.
In addition, the inclusion of $\gamma$ automatically promotes plane-wave solutions that are sparse in the set of plane-wave coefficients. To see this, consider $\hat{\alpha}_\ell$, the $\ell$th element of $\hat{\bPhi}$. Then, with $\mu^{(0)}_\ell$ being the corresponding measure and $m_\ell^{(q)}$ the transport plan between the barycenter and the $q$th microphone,
\begin{align*}
    \int_{\TT\times\TT} \gamma dm_\ell^{(q)}(\psi_1,\psi_2) = \gamma\int_\TT d\mu_\ell^{(0)}(\psi) \geq \gamma\abs{\int_\TT e^{i\psi} d\mu_\ell^{(0)}(\psi)}
\end{align*}
with equality if and only if $\mu_\ell^{(0)}$ has support in a single point. Thus, $\gamma$ also induces an $\ell_1$-penalty on $\hat{\bPhi}$, promoting sparse solutions.

It may be noted that the problem defining the proposed estimator is convex and amenable to off-the-shelf solvers.
\section{Numerical results}
\label{sec:num_res}
In this section, we evaluate the performance of the proposed methods on simulated examples in 2D. In the simulations, three plane-waves with frequency $f = 1$kHz impinge on a circular microphone array of radius 0.25m. Speed of sound is set to $\soundspeed = 343$m/s. In each simulation, the incident angles are picked uniformly at random on $[-\pi,\pi)$, with amplitudes drawn from a standard normal distribution and phases uniform on $[-\pi,\pi)$. The additive noise is drawn from a complex Gaussian distribution as $\epsilon^{(q)} \sim \mathcal{CN}(0,\sigma_\epsilon^2$), and the phase-perturbations are drawn as $\Delta_\ell^{(q)} \sim \mathcal{N}(0,\sigma_\Delta^2)$, approximating a von Mises distribution. Then, given noisy observations \eqref{eq:noisy_measurements}, we use the proposed method to recover $\bPhi$. As comparison, we also compute estimates of $\bPhi$ using the Tikhonov, Lasso, and LAD-Lasso estimators \cite{OkamotoLadlasso}. The proposed estimator as well as the comparison methods all use a common grid of $L = 50$ plane-waves with directions uniform on $[-\pi, \pi)$. For the proposed method, we discretize the transport problem in \eqref{eq:basic_ot} by gridding the phase-space $\TT$ into 50 points. Values for the hyper parameters for all methods are set using cross-validation \cite{berrar2019cross}. All estimators are implemented using the CVX library~\cite{grant2014cvx}. Three sets of simulations are performed. In the first case, the number of microphones is fixed to $Q = 9$, the plane-wave frequency is set to $f = 1\,\text{kHz}$, and the standard deviation of the phase perturbations, $\sigma_\Delta$, is varied. In the second case, $\sigma_\Delta = 0.5$~radians and $f = 1\,\text{kHz}$ are fixed, while the number of microphones $Q$ is varied. In the third case, the frequency is varied while $Q = 9$ and $\sigma_\Delta = 0.5$ are kept fixed. In all three scenarios, the noise variance $\sigma_\epsilon^2$ is set as to yield a fixed signal-to-noise ratio of $15$dB. As a measure of accuracy, we use the normalized mean-squared error (NMSE), the same metric as used in \cite{lluis2020sound}, which is defined as
\begin{equation*}
    \mathrm{NMSE}
    = \frac{\sum_{n=1}^{N} \abs{\langle g(\br^{(n)}), \hat{\bPhi}\rangle -\langle g(\br^{(n)}), \bPhi\rangle}^2}
           {\sum_{n=1}^{N} \abs{\langle g(\br^{(n)}), \bPhi\rangle}^2},
\end{equation*}
where $\br^{(n)}$, $n = 1,\ldots,N$, with $N = 3.6\times 10^4$ is a set of test points uniform on a $0.6\text{m}\times 0.6\text{m}$ domain. An illustration of the setup can be seen in Figure~\ref{NR:SFR}, together with the ground truth sound-field and estimates produced by the different methods in an example with $Q = 9$ microphones and $\sigma_\Delta = 0.5$ radians with $f=1.5$kHz. The same figure displays the magnitude of the estimated plane-wave coefficients. As can be seen, the proposed method gives a (visually) more accurate recovery of both the set of plane-wave coefficients and the sound-field. The results, averaged over 30 Monte Carlo simulations, for the experiments varying the  phase-perturbation deviation, number of microphones and frequency of the phase-waves are shown in Figures~\ref{fig:vary_parameters} left, middle and right, respectively. It may be particularly noted that the performance gap between the proposed method and the baselines becomes increasingly pronounced as more microphone measurements are added. The proposed method also performs better in low frequency estimation scenarios.
\section{Conclusions}
\label{sec:conclusion}
We presented a new method for ill-posed sound field estimation under errors-in-variables phase perturbations, using an optimal transport barycenter formulation. Observation-dependent plane-wave coefficients are lifted to non-negative measures on the phase circle and transported to a generalized average, barycenter. The proposed method corrected systematic phase shifts and enabled consistent coefficient recovery via inverse lifting, i.e., a first-Fourier mapping. A sparsity penalty on the transported mass regularizes the barycenter toward a Dirac-like measure and enhances stability with large plane-wave dictionaries, while preserving convexity. Simulations show improved accuracy and robustness to phase errors over baseline methods, even with additive noise. Future work will validate the approach on real measurements and extend it to broadband fields and spatially varying calibration errors.

\clearpage
\balance
\small
\bibliographystyle{IEEEbib}
\bibliography{refs.bib}

@article{gabriel2019computational,
  title={Computational Optimal Transport with Applications to Data Sciences},
  author={Gabriel, Peyr{\'e} and Marco, Cuturi},
  journal={Foundations and Trends{\textregistered} in Machine Learning},
  volume={11},
  number={5-6},
  pages={355--607},
  year={2019},
  publisher={Emerald Publishing Limited}
}

@book{villani2009optimal,
  title        = {Optimal Transport: Old and New},
  author       = {Villani, Cédric},
  series       = {Grundlehren der mathematischen Wissenschaften},
  volume       = {338},
  year         = {2009},
  publisher    = {Springer},
  address      = {Berlin, Heidelberg},
}

@ARTICLE{Kolouri2017omtApp,
  author={Kolouri, Soheil and Park, Se Rim and Thorpe, Matthew and Slepcev, Dejan and Rohde, Gustavo K.},
  journal={IEEE Signal Processing Magazine}, 
  title={Optimal Mass Transport: Signal processing and machine-learning applications}, 
  year={2017},
  volume={34},
  number={4},
  pages={43-59},
  doi={10.1109/MSP.2017.2695801}}

@article{RIRotBarycenter,
  title={Room Impulse Response Estimation through Optimal Mass Transport Barycenters},
  author={Pallewela, Rumeshika and Liu, Yuyang and Elvander, Filip},
  journal={arXiv preprint arXiv:2503.14207},
  year={2025}
}

@ARTICLE{WassersteinFourierD,
  author={Cazelles, Elsa and Robert, Arnaud and Tobar, Felipe},
  journal={IEEE Transactions on Signal Processing}, 
  title={The Wasserstein-Fourier Distance for Stationary Time Series}, 
  year={2021},
  volume={69},
  number={},
  pages={709-721},
  doi={10.1109/TSP.2020.3046227}}

@ARTICLE{OMToeplitzInterExtra,
  author={Elvander, Filip and Jakobsson, Andreas and Karlsson, Johan},
  journal={IEEE Transactions on Signal Processing}, 
  title={Interpolation and Extrapolation of Toeplitz Matrices via Optimal Mass Transport}, 
  year={2018},
  volume={66},
  number={20},
  pages={5285-5298},
  doi={10.1109/TSP.2018.2866432}}

@inproceedings{crocco2015room,
  title={Room impulse response estimation by iterative weighted $\ell$1-norm},
  author={Crocco, Marco and Del Bue, Alessio},
  booktitle={2015 23rd European Signal Processing Conference (EUSIPCO)},
  pages={1895--1899},
  year={2015},
  organization={IEEE}
}

@article{mignot2013low,
  title={Low frequency interpolation of room impulse responses using compressed sensing},
  author={Mignot, R{\'e}mi and Chardon, Gilles and Daudet, Laurent},
  journal={IEEE/ACM Transactions on Audio, Speech, and Language Processing},
  volume={22},
  number={1},
  pages={205--216},
  year={2013},
  publisher={IEEE}
}

@INPROCEEDINGS{koyamaKernel,
  author={Ueno, Natsuki and Koyama, Shoichi and Saruwatari, Hiroshi},
  booktitle={2018 16th International Workshop on Acoustic Signal Enhancement (IWAENC)}, 
  title={Kernel Ridge Regression with Constraint of {H}elmholtz Equation for Sound Field Interpolation}, 
  year={2018},
  volume={},
  number={},
  pages={1-440},
  doi={10.1109/IWAENC.2018.8521334}}

@INPROCEEDINGS{Davidkernel,
  author={Sundström, David and Koyama, Shoichi and Jakobsson, Andreas},
  booktitle={2024 18th International Workshop on Acoustic Signal Enhancement (IWAENC)}, 
  title={Sound Field Estimation Using Deep Kernel Learning Regularized by the Wave Equation}, 
  year={2024},
  volume={},
  number={},
  pages={319-323},
  doi={10.1109/IWAENC61483.2024.10694575}}

@INPROCEEDINGS{JesperSZC,
  author={Brunnström, Jesper and van Waterschoot, Toon and Moonen, Marc},
  booktitle={2023 31st European Signal Processing Conference (EUSIPCO)}, 
  title={Sound Zone Control for Arbitrary Sound Field Reproduction Methods}, 
  year={2023},
  volume={},
  number={},
  pages={341-345},
  doi={10.23919/EUSIPCO58844.2023.10289995}}

@ARTICLE{koyamaSFC,
  author={Abe, Takumi and Koyama, Shoichi and Ueno, Natsuki and Saruwatari, Hiroshi},
  journal={IEEE/ACM Transactions on Audio, Speech, and Language Processing}, 
  title={Amplitude Matching for Multizone Sound Field Control}, 
  year={2023},
  volume={31},
  number={},
  pages={656-669},
  doi={10.1109/TASLP.2022.3231715}}

@inproceedings{nicol19993d,
  title={3D-sound reproduction over an extensive listening area: A hybrid method derived from holophony and ambisonic},
  author={Nicol, Rozenn and Emerit, Marc},
  booktitle={Audio Engineering Society Conference: 16th International Conference: Spatial Sound Reproduction},
  year={1999},
  organization={Audio Engineering Society}
}

@article{hulsebos2002improved,
  title={Improved microphone array configurations for auralization of sound fields by wave-field synthesis},
  author={Hulsebos, Edo and de Vries, Diemer and Bourdillat, Emmanuelle},
  journal={Journal of the Audio Engineering Society},
  volume={50},
  number={10},
  pages={779--790},
  year={2002},
  publisher={Audio Engineering Society}
}

@article{koyama2021spatial,
  title={Spatial active noise control based on kernel interpolation of sound field},
  author={Koyama, Shoichi and Brunnstr{\"o}m, Jesper and Ito, Hayato and Ueno, Natsuki and Saruwatari, Hiroshi},
  journal={IEEE/ACM Transactions on Audio, Speech, and Language Processing},
  volume={29},
  pages={3052--3063},
  year={2021},
  publisher={IEEE}
}

@ARTICLE{samarasinghe2016noisecontrol,
  author={Samarasinghe, Prasanga N. and Zhang, Wen and Abhayapala, Thushara D.},
  journal={IEEE Signal Processing Magazine}, 
  title={Recent Advances in Active Noise Control Inside Automobile Cabins: Toward quieter cars}, 
  year={2016},
  volume={33},
  number={6},
  pages={61-73},
  doi={10.1109/MSP.2016.2601942}}

@article{betlehem2005theory,
  title={Theory and design of sound field reproduction in reverberant rooms},
  author={Betlehem, Terence and Abhayapala, Thushara D},
  journal={The Journal of the Acoustical Society of America},
  volume={117},
  number={4},
  pages={2100--2111},
  year={2005},
  publisher={Acoustical Society of America}
}

@article{samarasinghe2015efficient,
  title={An efficient parameterization of the room transfer function},
  author={Samarasinghe, Prasanga and Abhayapala, Thushara and Poletti, Mark and Betlehem, Terence},
  journal={IEEE/ACM Transactions on Audio, Speech, and Language Processing},
  volume={23},
  number={12},
  pages={2217--2227},
  year={2015},
  publisher={IEEE}
}

@article{wu2000reconstruction,
  title={On reconstruction of acoustic pressure fields using the {H}elmholtz equation least squares method},
  author={Wu, Sean F},
  journal={The Journal of the Acoustical Society of America},
  volume={107},
  number={5},
  pages={2511--2522},
  year={2000},
  publisher={Acoustical Society of America}
}

@article{buck2002aspects,
  title={Aspects of first-order differential microphone arrays in the presence of sensor imperfections},
  author={Buck, Markus},
  journal={European transactions on telecommunications},
  volume={13},
  number={2},
  pages={115--122},
  year={2002},
  publisher={Wiley Online Library}
}

@article{verburg2018reconstruction,
  title={Reconstruction of the sound field in a room using compressive sensing},
  author={Verburg, Samuel A and Fernandez-Grande, Efren},
  journal={The Journal of the Acoustical Society of America},
  volume={143},
  number={6},
  pages={3770--3779},
  year={2018},
  publisher={AIP Publishing}
}

@ARTICLE{georgios2010lasso,
  author={Lilis, Georgios N. and Angelosante, Daniele and Giannakis, Georgios B.},
  journal={IEEE Transactions on Audio, Speech, and Language Processing}, 
  title={Sound Field Reproduction using the Lasso}, 
  year={2010},
  volume={18},
  number={8},
  pages={1902-1912},
  doi={10.1109/TASL.2010.2040523}}

@INPROCEEDINGS{OkamotoLadlasso,
  author={Okamoto, Takuma},
  booktitle={2025 IEEE Workshop on Applications of Signal Processing to Audio and Acoustics (WASPAA)}, 
  title={SFC-L1: Sound Field Control With Least Absolute Deviation Regression}, 
  year={2025},
  volume={},
  number={},
  pages={1-4},
  doi={10.1109/WASPAA66052.2025.11230967}}

@article{lluis2020sound,
  title={Sound field reconstruction in rooms: Inpainting meets super-resolution},
  author={Lluis, Francesc and Martinez-Nuevo, Pablo and Bo M{\o}ller, Martin and Ewan Shepstone, Sven},
  journal={The Journal of the Acoustical Society of America},
  volume={148},
  number={2},
  pages={649--659},
  year={2020},
  publisher={AIP Publishing}
}

@incollection{berrar2019cross,
  title={Cross-validation.},
  author={Berrar, Daniel},
  publisher = {Elsevier Inc},
  booktitle = {Encyclopedia of Bioinformatics and Computational Biology},
  pages = {542-545},
  year={2019}
}

@article{golub1980analysis,
  title={An analysis of the total least squares problem},
  author={Golub, Gene H and Van Loan, Charles F},
  journal={SIAM journal on numerical analysis},
  volume={17},
  number={6},
  pages={883--893},
  year={1980},
  publisher={SIAM}
}

@ARTICLE{BetlehemPSZ,
  author={Betlehem, Terence and Zhang, Wen and Poletti, Mark A. and Abhayapala, Thushara D.},
  journal={IEEE Signal Processing Magazine}, 
  title={Personal Sound Zones: Delivering interface-free audio to multiple listeners}, 
  year={2015},
  volume={32},
  number={2},
  pages={81-91},
  doi={10.1109/MSP.2014.2360707}}

@book{williams1999fourier,
  title={Fourier acoustics: sound radiation and nearfield acoustical holography},
  author={Williams, Earl G},
  year={1999},
  publisher={Elsevier}
}

@book{kuttruff2016room,
  title={Room acoustics},
  author={Kuttruff, Heinrich},
  year={2016},
  publisher={Crc Press}
}

@article{antonello2017room,
  title={Room impulse response interpolation using a sparse spatio-temporal representation of the sound field},
  author={Antonello, Niccolo and De Sena, Enzo and Moonen, Marc and Naylor, Patrick A and Van Waterschoot, Toon},
  journal={IEEE/ACM Transactions on Audio, Speech, and Language Processing},
  volume={25},
  number={10},
  pages={1929--1941},
  year={2017},
  publisher={IEEE}
}

@article{royster2003sound,
  title={Sound Measurement: Instrumentation and Noise},
  author={Royster, L and Royster, J and Driscoll, D and Layne, M},
  journal={Noise Manual},
  volume={41},
  year={2003}
}

@ARTICLE{PSMOT,
  author={Georgiou, Tryphon T. and Karlsson, Johan and Takyar, Mir Shahrouz},
  journal={IEEE Transactions on Signal Processing}, 
  title={Metrics for Power Spectra: An Axiomatic Approach}, 
  year={2009},
  volume={57},
  number={3},
  pages={859-867},
  doi={10.1109/TSP.2008.2010009}}

@ARTICLE{RIRInterDavid,
  author={Sundström, David and Elvander, Filip and Jakobsson, Andreas},
  journal={IEEE Transactions on Signal Processing}, 
  title={Optimal Transport Based Impulse Response Interpolation in the Presence of Calibration Errors}, 
  year={2024},
  volume={72},
  number={},
  pages={1548-1559},
  doi={10.1109/TSP.2024.3372249}}

@ARTICLE{RIRSimuInfe,
  author={Björkman, Anton and Sundström, David and Jakobsson, Andreas and Elvander, Filip},
  journal={IEEE Transactions on Signal Processing}, 
  title={Optimal Transport Regularization for Simulation-Informed Room Impulse Response Estimation}, 
  year={2025},
  volume={73},
  number={},
  pages={5244-5256},
  doi={10.1109/TSP.2025.3643595}}

@article{sundstrom2025boundary,
  title={Boundary-Informed Sound Field Reconstruction},
  author={Sundstr{\"o}m, David and Elvander, Filip and Jakobsson, Andreas},
  journal={arXiv preprint arXiv:2506.13279},
  year={2025}
}

@misc{grant2014cvx,
  title={{CVX}: Matlab software for disciplined convex programming, version 2.1},
  author={Grant, Michael and Boyd, Stephen},
  year={2014}
}

\end{document}